\documentclass[letter]{aa} 
\pdfoutput=1                                   
\usepackage{graphicx}
\usepackage{hyperref}
\usepackage{txfonts}
\usepackage{bm}
\usepackage{natbib}
\begin{document}
%
\title{ Simulation of dark lanes in post--flare supra--arcades}
  \subtitle{III. A 2D simulation}

   \author{L. S. Maglione
          \inst{1}
          \and
         E. M.  Schneiter \inst{2}
	\and A. Costa \inst{2,3,4} \fnmsep\thanks{CA: Andrea Costa, acosta@mail.oac.uncor.edu}
	\and S. Elaskar \inst{3,4} 
          }

   \institute{Facultad de Ingenier\'\i a, Universidad Nacional de R\'\i o Cuarto, Argentina  
                  \and
     Instituto de Astronom\'\i a Te\'orica y Experimental, C\'ordoba, Argentina 
          \and      Facultad de Ciencias Exactas, F\'\i sicas  y Naturales, Universidad Nacional de C\'ordoba, Argentina 
     \and      Consejo Nacional de Investigaciones Cient\'\i ficas y T\'ecnicas 
}
   \date{Received ; accepted }

 
  \abstract {Using two simulations of $1.5D,$ for the first time, in
    Costa et al. (2009) and Shulz et al. (2010) we numerically
    reproduce the observational dark inflows described in Verwichte et
    al. (2005).  We show that the dark tracks can be explained as hot
    plasma vacuums generated upstream of a slow magnetoacoustic shock
    wave produced by a localized deposition of energy.  } {To confirm
    the later `dark lane' interpretation and to identify specific $2D$
    contributions to the description of the phenomenon. } {To solve
    the ideal and non--stationary MHD equations we used a $2D$ Riemann
    solver Eulerian code specially designed to capture supersonic flow
    discontinuities. } { The numerical $2D$ results are in agreement
    with the observational behaviour however they show a slight shift
    in the characteristic parameter with respect to those found
    previously. } {We confirm qualitatively the behaviour found in the
    previous papers.  For a given numerical domain the period of the
    kink--like structure is a function of the magnetic field
    intensity: larger periods are associated with lower magnetic field
    intensities.  Contrary to the $1D$ result -where the sunward
    dynamic is independent of the magnetic field intensity due to its
    exclusively waveguide role- in the $2D$ simulation the sunward
    speed is larger for larger values of the magnetic field.  This can
    be interpreted as the capability of the low coronal plasma to
    collimate the deposition of energy into the magnetic field
    direction.  The moving features consistent of low--density and
    high--temperature plasma cavities have larger inside values of the
    structuring parameter $\beta$ than the neighboring media.  Thus,
    the voids seem to be the emergence structures of a whole nonlinear
    interacting plasma context of shocks and waves more than voided
    plasma loops magnetically structured. }


   \maketitle
%

\section{Introduction}
Dark sunward sinuous lanes moving along a fan of rays above
post--flare loops towards a supra--arcade have been extensively
studied (Innes et al. \citealp{inna}a; \citealp{innb}b; McKenzie and
Savage \citealp{mck3}).  The down moving structures observed at
$[40-60]$Mm heights above the top of arcades, with a decelerating
speed in the range $\sim [50 - 500] $kms$^{-1}$ were interpreted as
sunward voided flows generated by reconnection processes developed by
a current sheet above the flare arcade. Another configuration
consistent with the observations is a magnetic flux tube, filled with
flux and very little plasma, shrinking into the post-eruption arcade
(McKenzie and Hudson \citealp{mck}; McKenzie \citealp{mck2}). Recently
Linton et al. \cite{lin}, proposed another description where the
dynamic is triggered by a localized reconnection event that produces
up and down flowing reconnected flux tubes which decelerate due to a
underlying magnetic arcade loops.  Verwichte et al. (\citealp{ver},
hereafter VNC) analyzed transverse to the magnetic field oscillations
associated with sunward dark lanes in a post--flare
supra--arcade. They found that the initial speeds and the displacement
amplitudes, of observational dark lanes of a kink--like type, decrease
as they propagate downwards while the period remains constant with
height. 
increase.
 
In Costa et al. \cite{pap1} and in Schulz et al. \cite{sch} (hereafter
Paper 1 and 2, respectively), by the integration of two $1.5D$ MHD
ideal equations, we presented a new scenario that gives account of the
observational dark voids described in VNC.  We simulated the effects
of an initial impulsive and localized deposition of energy -supposed
to be associated with above reconnection processes- in a plasma
structured by sunward magnetic field lines.  The impulsive phase was
modeled by a pressure perturbation that initiates two main processes,
a fundamentally hydrodynamic shock pattern directed sunwards and a
perpendicular magnetic shock one, i.e., transverse to the magnetic
field. The two patterns were supposed to be independent processes,
however linked by their common origin and background magnetic and
density conditions. The independence of the two dynamics was justified
due to the far more effective conductive energy transport along field
lines than across them.

   We showed that the dark tracks are consistent with plasma voids
   generated by the bouncing and interfering of shocks and expansion
   waves upstream the initial localized deposition of energy, which is
   responsible of the two dynamics.  The composition of both, a
   resulting sunward directed hydrodynamic shock pattern and a
   perpendicular magnetic shock one, produces an overall $\beta>1$
   transversely shaking void moving towards the sun surface, that
   resembles the kink--like mode described in VNC.
 
  From the transverse simulation we found that, in accordance with the
  shock wave theory of uniform media (Kirk et al. \citealp{kirk}),
  there is a critical value of the magnetic field beyond which the
  behaviour of the magnetic shock pattern changes, the magnetic
  compression is limited and the phenomenon is progressively
  saturated.  We also found that the period of the kink--like
  structure is a function of the magnetic field intensity while the
  amplitude is a function of the triggering pressure pulse.  Thus, the
  period's constancy with height found in VNC can be associated with
  the almost constancy with height of the background magnetic field in
  the region.  In accordance with our interpretation that in the
  sunward direction the magnetic field plays the role of being a
  wave--guide, we found that the pressure pulse determines an
  hydrodynamic shock that moves towards the sun surface at slow
  acoustic speed values and the dynamic is independent of the magnetic
  field intensity.
    
 In this paper we investigate the goodness of the 1D model and we
 discuss the limitations and new characteristics of the phenomenon
 associated to the 2D structure.

\section{Numerical Code and Initial Conditions}   

We integrate the MHD equations using the two-dimensional version of
the `Mezcal' code, an Eulerian Godunov MHD code (De Colle and Raga
\citealp{dec1}, \citealp{dec1}, De Colle and al. \citealp{dec3}).  The
MHD Riemann solver uses a standard second-order Runge-Kutta method for
the time integration and a spatially second-order reconstruction of
the primitive variables at the interfaces (except in shocks). The
constrained transport method (e.g., T\'oth \citealp{toth}) is used to
conserve $\nabla \cdot \vec B = 0$ to machine accuracy. The code has
been extensively tested against standard problems (De Colle
\citealp{dec4}).
  
All calculations were performed using a numerical grid of
$(x,y)=(1600,620)$ grid--points corresponding to a physical size of
$(40,9)$Mm, chosen in accordance with VNC observations. The coordinate
$x$ represents the sunward direction and the $y$ coordinate the
transverse to the magnetic field one, as seen by the line of sight. We
assumed a constant radial initial magnetic field structure, a typical
background temperature of $T=3.0\times 10^{6}$K and a numerical
density value of $\rho=0.46\times 10^{9}$~cm$^{-3}$.  A spherical
(circular in 2D) relative triggering pressure pulse $P_{2}/P_{1}$
($P_{2}$ is the triggering pressure pulse and $P_{1}$ is the
background gas pressure of the corona) of radius $0.6$Mm was localized
in the position $(1520,434)$ equivalent to $(38,6.3)$Mm.

 Several simulation were carried out, varying the radial (normal to
 the sun's surface) magnetic field, $B ,$ and the localized deposition
 of energy, modelled as the relative triggering pressure pulse.  As in
 Paper 1 and 2, the boundary conditions were fixed, allowing rebounds
 in the lateral transverse directions and also in the upper radial
 direction, resembling the action of the reconnection site. We assumed
 that the perturbations are absorbed in the sunward direction.  The
 characteristic parameters were chosen in accordance with typical
 observed dark lane structures, e.g., VNC data and McKenzie and Hudson
 \cite{mck}.

\section{Results and Discussion}
\begin{figure}
\centering
  \includegraphics[width=6.cm]{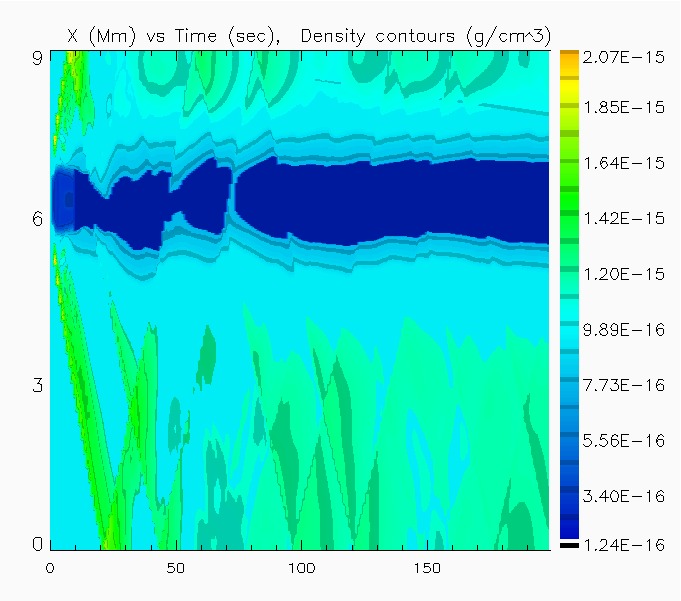}
  \includegraphics[width=6.cm]{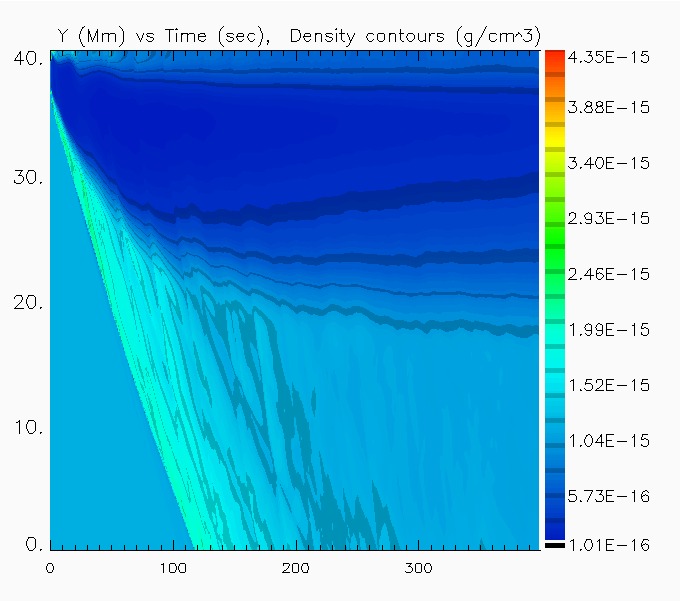}
   \includegraphics[width=6.cm]{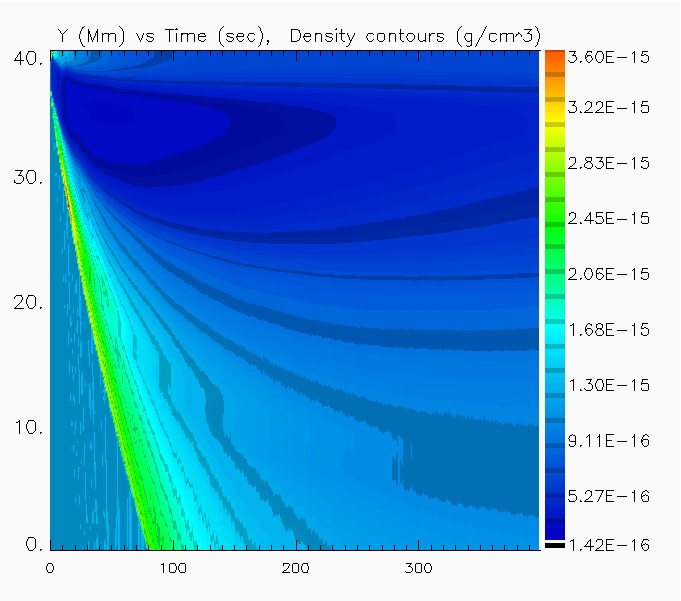}
  \caption{Density patterns ($g cm^{-3}$) a) transverse kink--like
    oscillation of the dark lane as a function of time for $4$G.  b)
    and c) sunward motion as a function of time for $4$G and $20$G,
    respectively.  $P_{2}/P_{1}=100$G.}
              \label{fig:uno}
    \end{figure}

Figures~\ref{fig:uno}a,b show the density pattern in the central
transverse direction and respectively, in the central longitudinal
direction, as a function of time for $P_{2}/P_{1}=100$ and $B=4$G. The
figures are qualitatively similar to the $1D$ simulations (see Paper 1
and 2).  The transverse kink--like structure is discontinued at a time
step $\sim 80$sec apparently due to a strong shock front that has
swept material locally erasing the vacuum trace. These features -that
remind observational constrictions, see e.g., Innes et al.
\cite{inna}a; \cite{innb}b- were found in several cases at different
positions and times, they are the result of nonlinear interactions
difficult to predict.  For further comparison we varied the background
magnetic field.  Table 1 shows the numerical periods ($\tau$),
amplitudes ($A$), the distances evacuated by the moving perturbations
($L_{v}$), and the initial sunward speeds ($V_{0}$) obtained for the
$2$G, $4$G, $8$G and $20$G cases and for $P_{2}/P_{1}=100.$ These
results are consistent with characteristic observational values given
in literature, e.g., VNC, McKenzie and Hudson \cite{mck}.  However, in
order to match periods of decades of seconds and amplitudes of
hundreds of kilometers we required pressure pulses of almost an order
of magnitude larger than in the $1D$ case.  As before, we found that
lower values of the magnetic field are associated with larger periods
(see second column of Table 1).  Also, the phenomenon is progressively
saturated resembling the compressional limits (in density and magnetic
field intensity) of the HD and transverse MHD shock wave theory in
uniform medium (Kirk et al. \citealp{kirk}), i.e., the augmentation of
the magnetic field from $4$G to $6$G implies a decrease of $55$sec in
the period while the change of magnetic field from $8$G to $20$G
implies a decrease of only $10$sec in the period. There is no visible
relation between the magnetic field and the amplitude.

 Figure~\ref{fig:uno}c represents the void density for the
 longitudinal dynamic of the $20$G magnetic field.  Contrary to the
 sunward behaviour found in the $1D$ simulations, when comparing
 Fig.~\ref{fig:uno}c with the $4$G case of Fig.~\ref{fig:uno}b, the
 void associated with larger magnetic field intensities travels a
 larger distance in the sunward direction (see Table 1, fourth
 column).  In Paper 2, the hypothesis of independence between the two
 $1D$ simulations was justified by the anisotropy imposed by the
 magnetic field, thus its role was solely to wave guide the
 longitudinal dynamic. In the $2D$ simulation the deposition of energy
 is distributed radially, from the location of the initial
 perturbation, to all directions of the $(x,y)$ plane. However, the
 freezing in of the plasma to the magnetic field induces the
 collimation of part of the energy towards the longitudinal
 direction. The more intense magnetic fields are the more efficient in
 line--tying the energy, thus, giving impulse to the final resulting
 sunward motion. Hence, the larger the magnetic field the more
 accurate the $1D$ description will be.

\begin{figure*}
  \centering
   \includegraphics[width=12.cm,height=12.cm]{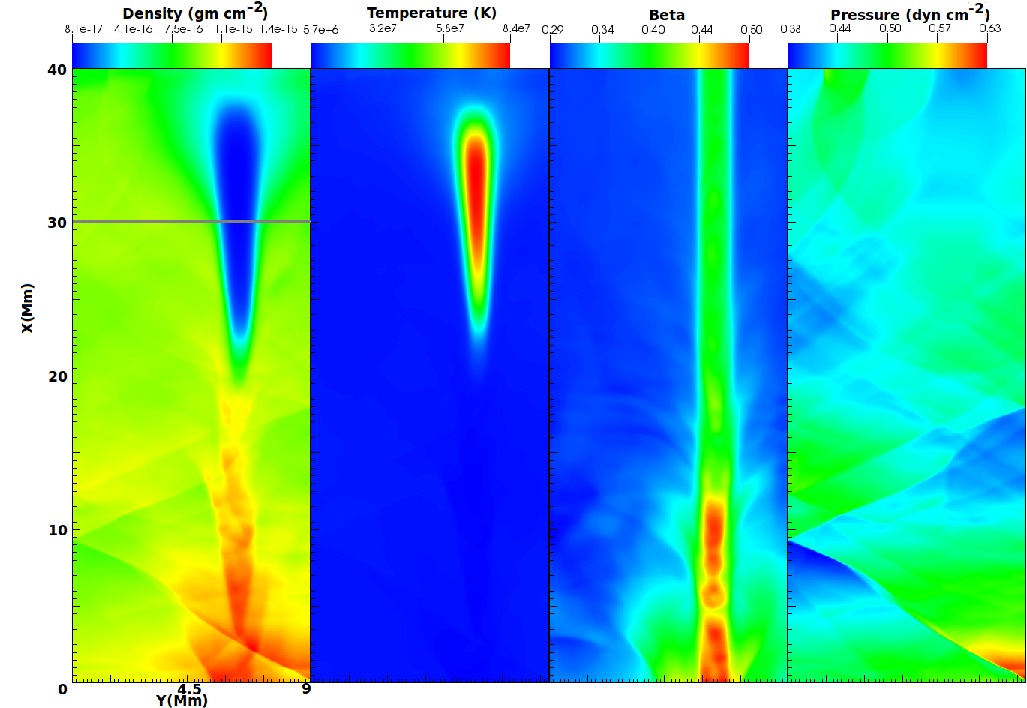}
     \caption{a) density, b) temperature, c) $\beta$ parameter and d)
       total pressure of the dark lane pattern. For all the cases we
       used the $(x,y)$ grid defined as in VNC, $x$ the radial
       direction and $y$ the transverse direction, the magnetic field
       is $4$G, the time is $200$sec, and the triggering pulse is
       $P_{2}/P_{1}=100$G.}
    \label{fig:dos}%
    \end{figure*}

\begin{figure}
   \centering
       \includegraphics[width=6.cm]{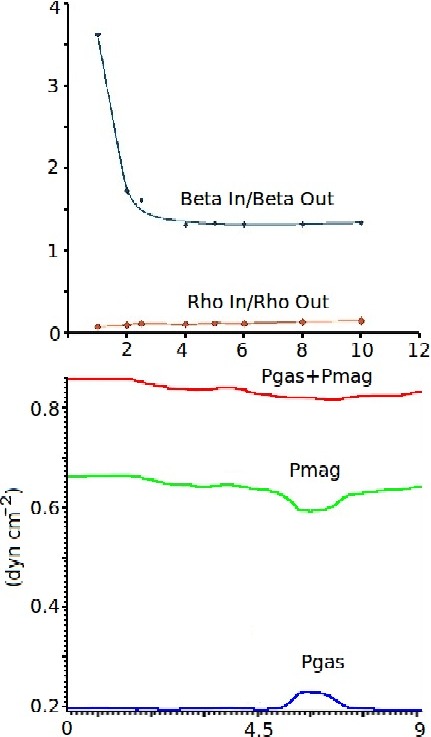}
    \caption{a) $\beta$ average inner to outer ratio values for
      different magnetic field intensity cases.  b) gas, magnetic and
      total pressure values for the grid positions of the line marked
      in Fig.~\ref{fig:dos}a and $B=4$G. For all cases the triggering
      pulse is $P_{2}/P_{1}=100$G and $\tau=200.$}
    \label{fig:tres}
    \end{figure}

Figure~\ref{fig:dos}a-d shows, respectively the `dark lane' density
structure, the temperature, the parameter $\beta$ and the total
pressure displayed in the whole numerical domain for $4$G and
$P_{2}/P_{1}=100$ at $t=200$sec.  As in Papers 2, in accordance with
observational data, the vacuum density is almost an order of magnitude
less than the external medium (see Fig.~\ref{fig:tres}a), while the
vacuum temperature is almost an order of magnitude higher than the
temperature in the surroundings.  When plotting $\beta$ we see an
extended channel (not confined to the vacuum zone) of inner $\beta$
values lower than the outside ones.  This behaviour occurs for the
other cases, with different magnetic field intensities.
Figure~\ref{fig:tres}a shows the ratio of an average inner to an
average outer $\beta$ value for the different cases studied at a fix
height of $35$Mm.  This indicates, as in Paper 2, that the inner
magnetic pressure is not responsible of preventing the collapse of the
vacuum zone.  Furthermore, from Fig.~\ref{fig:dos}d, it is possible to
see that the features that characterize the vacuum zone have
disappeared meaning that the contour of the cavity is, on average, in
total pressure equilibrium. This confirms that the void dynamics must
be sustained by the interaction of nonlinear waves and shocks acting
in times comparable to the observations. For a more detailed
description Fig.~\ref{fig:tres}b shows respectively, the gas, the
magnetic and the total pressure at each point corresponding to the
line traced in Fig.~\ref{fig:dos}a. Note that the almost constancy of
the total pressure is due to the gas pressure increase (which can only
be accomplished through large temperature values) accompanied of a
magnetic pressure decrease through the vacuum region.

In Paper 2 we found that the cases studied had $\beta > 1,$ whereas in
the $2D$ simulations only the cases with magnetic fields lower than
$3$G have $\beta >1.$ However, as in the $1D$ case, the behaviour is
explained by the fact that $\beta $ is larger inside than outside the
voided cavity, i.e., the cavity is a result of dynamic nonlinear
interacting waves and shocks sustaining the whole pattern.  Note the
differentiated shape of the Fig.~\ref{fig:tres}a curve. The almost
constant value of the curve for large values of the magnetic field
intensity indicate a uniform behaviour within this range. The steep
negative slope of large inner values of $\beta $ and small values of
the magnetic field, up to $B \sim 3$G, exhibit a dynamic where the
anisotropy imposed by the magnetic field is softened and $2D$ features
become more significant. However, magnetic field intensities lower
than $B \sim 2.5$G have non--perturbed $\beta >1$ values which are not
realistic parameters for the low corona description. We conclude that
the two $1D$ simulations give a good approximate description of the
phenomenon.

Four runs, with different $P_{2}/P_{1},$ were carried out to analyze
the behaviour of the voids.  Table 2 shows the periods $\tau$, the
amplitudes $A$, the distances evacuated by the perturbations $L_{v},$
and the initial speed for a magnetic field intensity of $8$G as a
function of $P_{2}/P_{1}.$ The obtained parameters are in
correspondence with typical observed parameter ranges (see Table 1 in
VNC). Note that from Table 2 we see that, as in Paper 2, the increase
of $P_{2}/P_{1},$ implies the increase of the initial speed and the
distance evacuated by the perturbation, whereas the periods and
amplitudes show no regular variation.


 \begin{table}
 \centering
\begin{tabular}{ccccc}
$B [G]$&$\tau[sec] $& $A[km]$&$L_{v}[Mm]$ &$V_{o}[km/sec]$\\
 \hline  
$4 $&$80$&$725$&$22$&$200$\\
 \hline
  $ 6$& $25$&$825$&$24$&$250$\\
 \hline
$ 8  $ & $20$&$825$&$26$&$259$\\
 \hline
 $ 20  $ & $10$&$750$&$40$&$280$\\
 \hline
 \hline
\end{tabular}
\caption{\label{tab:table1} Numerical $2D$ parameters. $\vec{B}$ the
  magnetic field in the radial direction, $\tau$ the period, $A$ the
  amplitude, $L_{v}$ the distance evacuated by the perturbations and
  $V_{o} $ the initial sunward speed.  $P_{2}/P_{1}=100$G.}
\end{table}
\begin{table}
 \centering
\begin{tabular}{cccccc}
$P_{2}/P_{1}$&$\tau[sec] $& $A[km]$&$L_{v}[Mm]$ &$V_{o}[km/sec]$\\
 \hline  
$50$&$20$&$375$&$12$&$140$ \\
 \hline
  $ 100$& $25$&$1000$&$24$&$259$ \\
 \hline
$ 150$ & $25$&$750$&$23$&$280$ \\
 \hline
 $ 200  $ & $ -- $ &$1875$&$32$&$318$ \\
 \hline
 \hline
\end{tabular}

\caption{\label{tab:table2} Numerical $2D$ parameters. $P_{2}/P_{1}$
  the triggering pressure pulse, $\tau$ the period, $A$ the amplitude,
  $V_{o} $ the initial sunward speed and $L_{v}$ the distance
  evacuated by the perturbations. For all cases $B=8$G.}
\end{table}
\section{Conclusions}
We performed a $2D$ numerical simulation to reproduce observed dark
sinuous lanes moving sunwardly towards supra--arcades.  In previous
papers we reproduce observational features of the phenomenon using two
$1D$ simulations -supposed to be independent due to the far more
effective transport along the field lines than across them- that were
triggered by a pressure pulse acting on the transverse and
longitudinal directions. The correspondent transverse and longitudinal
patterns obtained from the $2D$ simulation are in good agreement with
the observations, confirming that the two $1D$ description is suitable
for this phenomenon.

The features, travelling decades of Mm, are voided cavities that
elongate towards the sun decelerating at sounds speeds, i.e., at
hundreds of kms$^{-1}.$ The $2D$ results shown in Table 1 reproduce
typical periods, amplitudes, distances travelled by the perturbations
and their initial speeds. However, the pressure pulse required is
almost an order of magnitude higher than in the $1D$ simulations. This
seems reasonable due to the fact that the pulse must trigger the
dynamic in all the radial directions and not only in two main ones.
As shown in Table 2, the increase of the pressure pulse that triggers
the phenomenon augments the initial speed and the distance travelled
by the front of the voided cavity. Linton et al. \cite{lin} also
reproduce the sunward motion though they assume open lateral
boundaries and a bottom reconnection site.  A main difference with our
work is that the reconnection process is intrinsic to their model as
the voids are formed and confined by a current sheet. From their work
is not clear if an oscillating behaviour, as described in VNC, is
obtained.

As in earlier works, we found that lower values of the magnetic field
are associated with larger transverse periods.  Also, we find from the
behaviour of the period that, beyond a critical magnetic field value
(between $4 $G and $6$G), the phenomenon is progressively saturated
with increasing values of the magnetic field. This reminds the
compressional limits, in density and magnetic field intensity, of the
HD and MHD shock wave theory for uniform media. However, the
determination of the critical magnetic field value of this more
complex scenario might be conditioned by other parameters not taken
into account in this work such as the distance between the radial
boundaries and the radius and the shape of the perturbation.  The
freezing in of the plasma to the magnetic field induces the
collimation of part of the energy towards the sunward
direction. Hence, larger magnetic field intensities produce features
travelling farther.

 Due to the fact that $\beta$ is larger inside than outside the voided
 cavity, and that its contour is in total pressure equilibrium we
 conclude that the internal magnetic pressure cannot be responsible of
 preventing the collapse of the vacuum zone.
 
 The $2D$ simulation shows, as in recent papers, that the sunwardly
 moving shaking voids are produced and sustained by two main processes
 that can be viewed as almost independent phenomenon if the magnetic
 field intensity is sufficiently high: 1) the interaction of nonlinear
 waves and shocks that rebound in the lateral denser medium, and 2)
 the interaction of nonlinear waves and shocks that upwardly rebound
 and are absorbed sunwardly.

 


\begin{thebibliography}{}
\bibitem[2009]{pap1} Costa, A., Elaskar, S., Fern\'andez, C., Mart\'\i
  nez, G., MNRAS 400, L85, 2009
\bibitem[2005]{dec1} De Colle, F., Raga, A.C.,  MNRAS 359, 164, 2005
\bibitem[2006]{dec2} De Colle, F., Raga, A.C.,  A\&A 449, 1061, 2006
\bibitem[2008]{dec3} De Colle, F., Raga, A.C., Esquivel, A., ApJ 689, 302, 2008
\bibitem[2005]{dec4} De Colle F., 2005, PhD thesis, UNAM
\bibitem[2009]{fe09} Fern\'andez, C., Costa, A., Elaskar, S., Shulz,
  W., MNRAS, 400,1821,2009
\bibitem[2003]{inna} Innes, D.E., McKenzie, D., Wang, T.,
  Sol.Phys. 217, 247, 2003
\bibitem[2003]{innb} Innes, D.E., McKenzie, D., Wang, T.,
  Sol.Phys. 217, 267, 2003
\bibitem[1994]{kirk} Kirk, J., Melrose, D., Priest, E., `Plasma
  Astrophysics' (Berlin, Verlag), 1994
\bibitem[2009]{lin} Linton, M.G., DeVore, C.R., Longcope, D.W., Earth
  Planets Space, 61, 1, 2009
\bibitem[2000]{mck2} McKenzie, D., Sol.Phys. 195, 381, 2000
\bibitem[1999]{mck} McKenzie, D., Hudson, H., ApJ, 519, L93, 1999
\bibitem[2009]{mck3} McKenzie, D., Savage, S., ApJ 697, 1569, 2009
\bibitem[2010]{sch} Schulz, W., Costa, A., Elaskar, S., Cid, G.,
  MNRAS, 407, L89, 2010
\bibitem[2000]{toth} T\'oth G., 2000, Journal of Computational
  Physics, 161, 605
\bibitem[2005]{ver} Verwichte, E., Nakariakov, V., Cooper, F., A\&A
  430, L65, 2005
\end{thebibliography}
\end{document}